# Uncovering the roughness effect on inelastic phonon scattering and thermal conductance at interface via spectral energy exchange

Jinyuan Xu and Yangyu Guo[*]

*School of Energy Science and Engineering, Harbin Institute of Technology, Harbin 150001, China*

(Dated: Mar. 29, 2025)

**Abstract**

Understanding the mechanism of interfacial thermal transport is crucial for thermal management of electronics. Recent experiments have shown the strong impact of interfacial roughness on inelastic phonon scattering and interfacial thermal conductance (ITC), while the theoretical modeling and underlying physics remain missing. Through non-equilibrium molecular dynamics simulations with quantum correction, we predict ITC of both sharp and rough Si/Al interfaces in a good agreement with experimental results in a broad range of temperatures. We further introduce a novel spectral energy exchange analysis, which reveals more annihilation of high-frequency phonons and generation of moderate-frequency phonons around the sharp interface compared to its rough counterpart. However, the low-frequency phonons at rough interface shows unexpected stronger inelastic scattering and larger contribution to ITC due to unique emerging interfacial modes. Our work thus promotes both the methodology and understanding of interfacial thermal transport at solid/solid interfaces, and may benefit the design and optimization of thermal interface materials.

---

[*]yyguo@hit.edu.cn

## 1. Introduction

The ongoing miniaturization and integration of electronic systems has brought up substantial thermal management challenges, particularly regarding interfacial thermal resistance in micro- and nano-scale devices [1–5]. The microscopic understanding of interfacial thermal conductance (ITC) is a crucial state-of-art research focus, as evidenced by emerging experimental reports that directly observe localized phonon modes at the interfaces between dissimilar materials [6–8]. Phonons can cross the interface in an inelastic way, as been both theoretically [9–11] and experimentally [12–15] shown. The interfacial atomic details crucially govern phonon transport across realistic interfaces, including atomic-scale roughness [16–18], doping [19,20], intermediate transition layer [21,22], nano-structures [23] and so on [24]. It has also been demonstrated in experiments that the surface treatment method of the substrate plays a critical role in determining the ITC of the ultimately fabricated interface [25–28].

Classical theoretical models have been proposed including the acoustic mismatch model (AMM) [29] and diffuse mismatch model (DMM) [30] to compute interfacial thermal resistance. The AMM usually works well at very low temperatures due to its assumption of wave-like specular transmission of phonons [30–32]. The DMM assumes phonons as quasi-particles crossing the interface in a diffuse way, and captures the general trend of temperature-dependent ITC in experimental reports [12,30,33]. As a combination of AMM and DMM, mixed-type mismatch models have been further developed to discuss the interfacial roughness effect on ITC [34–37]. Nevertheless, both AMM and DMM have limited predictive accuracy due to two fundamental limitations: (*i*) the assumption of elastic scattering despite few efforts incorporating inelastic corrections [38,39], and (*ii*) the absence of consideration of interfacial atomic details. Moreover, these theoretical models are also difficult to uncover the underlying microscopic scattering mechanism.

Atomistic simulation capable of resolving interfacial atomic details have emerged as powerful approaches, particularly: (*i*) molecular dynamics (MD) simulations [9–11,40] that directly integrate classical atomic dynamics, and (*ii*) non-equilibrium Green's function (NEGF) formalism [41–44] that quantum mechanically addresses interfacial phonon transport. To have a deeper microscopic understanding, phonon spectral heat current (SHC) decomposition methodology [45] has been developed based on MD simulations in the pioneering works [11,46,47], and was later integrated into GPUMD package [48,49] to facilitate broader accessibility and usage. Previous studies have analyzed the role of inelastic phonon scattering at interface based on the anharmonic contribution to SHC extracted from MD simulation with the third-order interatomic force constants [11] or with an assumption of temperature-independent harmonic transmission function [50]. Moreover, there are further works to evaluate the inelastic contribution by the following measures via MD: (1) two metrics including the Kullback-Leibler divergence value and interfacial anharmonic ratio [51], (2) the difference between SHC in various positions along transport direction and that in two bulk materials [52]. Nevertheless, there is still limited understanding about how the inelastic phonon scattering occurs around the interface. Recently the local spectral energy exchange is introduced in the framework of NEGF method to show the net effect of inelastic scattering inside the interface region [44]. However, the NEGF method suffers from limitations in prohibitive computational costs especially for rough interfaces where large cross-section is required. In contrast, the MD simulations are capable of accounting for complex interfacial structures with significantly lower computational cost.

In this work, we introduce the spectral energy exchange in the framework of MD as a further step along the SHC analysis. With a careful choice of empirical potential and an inclusion of quantum correction, the predicted ITC of both sharp and rough Si/Al interfaces show a good agreement with the experimental results in a very broad range of temperatures. With increasing temperature, the phonon annihilation and generation are demonstrated to be stronger in the high-frequency and moderate-frequency ranges respectively around the sharp interface, which instead tend to saturate around the rough one. Moreover, the low-frequency phonons below ~3 THz shows unexpected inelastic scattering in the rough configuration due to special interfacial phonon modes unveiled by local vibrational density of states (VDOS).

Our work will contribute to both the methodology and deeper understanding of interfacial thermal transport. The remaining of this manuscript is organized as follows. The physical model and methodology will be introduced in Sec. 2. The results and discussion are presented in Sec. 3. Finally, Sec. 4 provides the concluding remarks.

**2. Methodology**

The interfacial thermal transport at Si/Al interface around various temperatures in this work is studied based on non-equilibrium molecular dynamics (NEMD) simulations using LAMMPS [53,54] package with a time step of 1 fs. The angular-dependent potential (ADP) [55] is used to describe the atomic interactions for Si-Si, Si-Al and Al-Al in both crystal and alloyed states, as it accurately describes the phonon dispersion compared to the classical modified embedded atom method (MEAM) potential [56] as detailed in Appendix A. The atomic configurations of sharp and rough Si/Al interface are presented in Fig. 1(a). Through systematic convergence tests (as detailed in Appendix B), we adopted a system length of 30 nm (thermal transport length 20 nm) and the cross-section of a 10 uc × 10 uc and 14 uc × 14 uc (uc = unit cell) for Si [111] and Al [111] sides, respectively, *i.e.*, cross-sectional area ~15.6 nm$^2$, for all subsequent simulations. To generate rough interface system, we firstly thermalize the interfacial region of sharp configuration at 3000 K for 0.5 ns under canonical (*NVT*) ensemble. The interfacial region includes the $L_0$ and $R_0$ regions as shown in Fig. 1(a), which extend 0.7 nm away from the interface position to both sides. Secondly, the interfacial region is cooled down from 3000 K to 80 K also under *NVT* ensemble with a cooling rate of 2 K/ps to generate interfacial amorphous $Si_xAl_{1-x}$ alloy. Then, the interfacial region is fully relaxed under *NVT* ensemble at 80 K for 0.5 ns to release the residual stress. Finally, the whole system was fully relaxed under isothermal-isobaric (*NPT*) ensemble for 1 ns at 80 K to generate an initial rough Si/Al interface system. The rough configurations at other target temperatures are simulated based on this initial rough Si/Al interface. Note that the length (~1.4 nm) of disordered rough interfacial region is consistent with that in experimental observation [14].

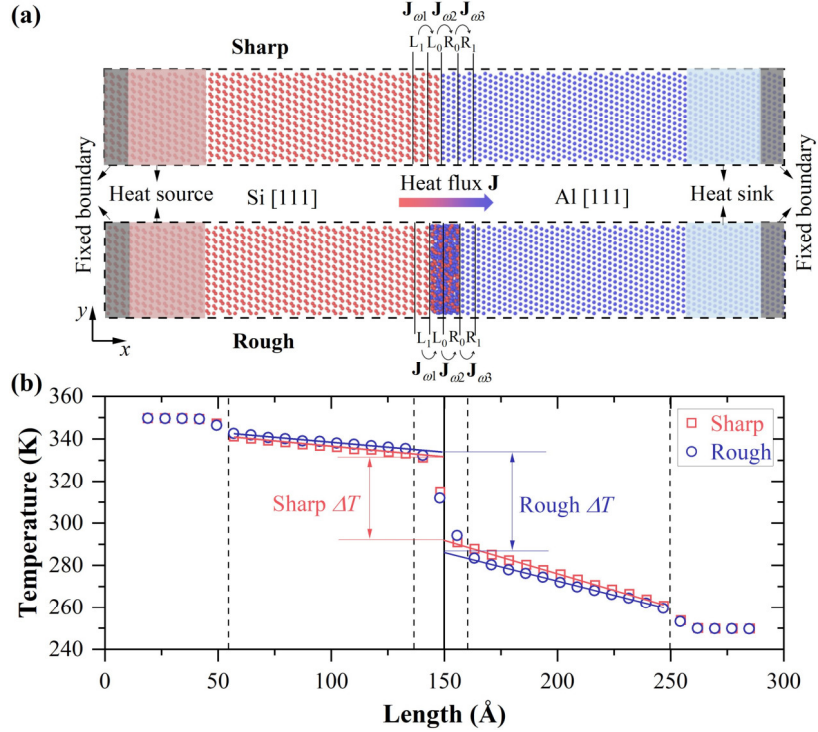

FIG. 1. MD simulation methodology: (a) Schematic of NEMD configurations of the atomically sharp and rough Si/Al interfaces. The $J_{\omega 1}$, $J_{\omega 2}$ and $J_{\omega 3}$ denote the spectral heat flux from $L_1$ to $L_0$ region, from $L_0$ to $R_0$ region and from $R_0$ to $R_1$ region, respectively, each region with a length of 7 Å. (b) The calculated steady state temperature profile along the thermal transport direction around a representative temperature of 300 K.

For each temperature, the whole system was fully relaxed under *NPT* ensemble at the target temperature for 1 ns. After structural optimization, the regions 1 nm away from both ends are fixed. The two 4 nm regions adjacent to the fixed ends are controlled by hot ($T_h$) and cold ($T_c$) Langevin thermostats under microcanonical (*NVE*) ensemble, respectively. The temperatures differences applied on the two thermostats are $T \pm 0.2T$ when $T < 300$ K and $T \pm 50$ K when $T > 300$ K to minimize nonlinear effect. The *NVE* simulations are carried out for 2 ns to achieve a steady state. Once the steady heat flux and temperature distribution are reached, another following 2 ns *NVE* data was used to calculate ITC. The interfacial temperature jump $\Delta T$ is extracted by linearly fitting the temperature distribution at two sides of the interface as shown in Fig. 1(b). Each ITC value was obtained through six independent NEMD simulations.

We collect the last 2 ns *NVE* data for SHC decomposition. The heat current from left contact to right contact $Q_{L \to R}$ can be decomposed spectrally as $Q_{L \to R} = \int_0^\infty q_{L \to R}(\omega) d\omega / 2\pi$, with $q_{L \to R}(\omega)$ given by [11,47,48]:

$$q_{L \to R}(\omega) = 2\operatorname{Re}\left[\int_{-\infty}^{\infty} -\sum_{i \in L}\sum_{j \in R} \left\langle \frac{\partial U_i}{\partial \mathbf{r}_j}(0) \cdot \mathbf{v}_j(t) - \frac{\partial U_j}{\partial \mathbf{r}_i}(0) \cdot \mathbf{v}_i(t) \right\rangle e^{i\omega t} dt \right], \quad (1)$$

where $\omega$ is the phonon angular frequency, and the atomic position, velocities and local potential energy of atom $i$ ($j$) are denoted by $\mathbf{r}_{i(j)}$, $\mathbf{v}_{i(j)}$ and $U_{i(j)}$, respectively. As the classical statistics of phonons are considered in MD simulations [57], which assumes that all phonon modes are equally activated even at very low temperatures. Hence, the quantum correction is introduced to the SHC as [41,58]:

$$q_{L \to R}^{QC}(\omega) = q_{L \to R}(\omega) \frac{\hbar \omega}{k_B} \frac{\partial f_{BE}}{\partial T}, \quad (2)$$

where $f_{BE}$ is the Bose-Einstein distribution, $\hbar$ is the reduced Planck's constant, and $k_B$ is the Boltzmann constant. The spectral heat flux, as denoted in Fig. 1(a), is computed by $\mathbf{J}_\omega = q_{L \to R}(\omega)/A$, where $A$ is the cross-sectional area.

The spectral energy exchange in NEGF formalism has been defined by [44]:

$$\delta \mathbf{J}_\omega = \frac{\hbar \omega}{2\pi} \frac{1}{A \cdot N} \sum_\mathbf{k} \operatorname{Tr}[\boldsymbol{\Sigma}_s^>(\omega; \mathbf{k})\mathbf{G}^<(\omega; \mathbf{k}) - \boldsymbol{\Sigma}_s^<(\omega; \mathbf{k})\mathbf{G}^>(\omega; \mathbf{k})], \quad (3)$$

where $N$ is the number of transverse wave vectors ($\mathbf{k}$) and 'Tr' denotes the trace of a square matrix. As $i\mathbf{G}^<$ and $i\boldsymbol{\Sigma}_s^>$ (i being the imaginary index) represents the phonon occupation and out-scattering rate, $-\boldsymbol{\Sigma}_s^> \mathbf{G}^<$ denote net phonon annihilation in the current state. Similarly, $-\boldsymbol{\Sigma}_s^<(\omega)\mathbf{G}^>(\omega)$ denotes net phonon generation in the current state. Hence, $\delta \mathbf{J}_\omega > 0$ and $\delta \mathbf{J}_\omega < 0$ refer respectively to net phonon generation and annihilation at a specific frequency $\omega$ due to inelastic phonon scattering. According to the energy balance relation, $\delta \mathbf{J}_\omega$ is also equal to the difference between the heat flux from the interface region to right side $\mathbf{J}_{\omega 2}$ and the heat flux from left side to the interface region $\mathbf{J}_{\omega 1}$, *i.e.*, $\delta \mathbf{J}_\omega = \mathbf{J}_{\omega 2} - \mathbf{J}_{\omega 1}$, as numerically verified in NEGF simulation [44]. Inspired by the NEGF formalism, we introduce the concept of spectral energy exchange around the interface in the framework of MD. The spectral thermal conductance is further calculated by $G_{\omega 1(2, 3)} = \mathbf{J}_{\omega 1(2, 3)} / \Delta T$, and the spectral energy exchange is thus defined as $\delta G_{\omega 12}$ and $\delta G_{\omega 23}$. To be more specific, $\delta G_{\omega 12} = G_{\omega 2} - G_{\omega 1}$ and $\delta G_{\omega 23} = G_{\omega 3} - G_{\omega 2}$ represent the net effect of inelastic phonon scattering in $L_0$ and $R_0$ regions shown in Fig. 1(a), respectively. Therefore, the spectral energy exchange provides some understanding about how inelastic scattering occurs around the interfaces. The local VDOS $\rho(\omega)$ is obtained by performing a Fourier transform of the velocity autocorrelation function [59]:

$$\rho(\omega) = \int_{-\infty}^{\infty} \sum_j \left\langle \frac{\mathbf{v}_j(0) \cdot \mathbf{v}_j(t)}{\mathbf{v}_j(0) \cdot \mathbf{v}_j(0)} \right\rangle e^{i\omega t} dt . \tag{4}$$

## 3. Results and Discussions

In Sec. 3.1, we firstly discuss the predicted overall ITC of sharp and rough Si/Al interfaces with a comparison to experimental data. Then, the spectral interfacial thermal conductance is also analyzed to understand the ITC results. In Sec. 3.2, the spectral energy exchange and local VDOS around the interface are further discussed to uncover the underlying inelastic phonon scattering mechanism.

### 3.1 Overall and spectral interfacial thermal conductance

First, we discuss the ITC results of Si/Al interfaces in comparison to the experimental phonon ITC $G_{pp}$ [14]. Based on the ADP empirical potential, our direct NEMD results generally agree with the experimental results above room temperature, while largely overestimate the experimental one at low temperature (< 200 K) as shown in Fig. 2(a). The MD results using MEAM potential above 300 K by Li *et al.* [14] are more or less consistent with our results in the same temperature range. However, explicit SHC decomposition and thus quantum correction were missing [14]. With quantum correction to the spectral thermal conductance as described in Sec. 2, our calculated ITC of both sharp and rough interfaces agree very well with experimental measurements as shown in Fig. 2(b), especially at low temperatures. It is worthy to explain that the amorphous $Si_xAl_{1-x}$ alloyed interface cannot be stably generated at temperatures above 500 K probably due to the limitations of the ADP empirical potential. Hence, the highest temperature for rough configuration is 450 K in our simulation, which is sufficient since the observed ITC tends to saturate beyond ~400 K.

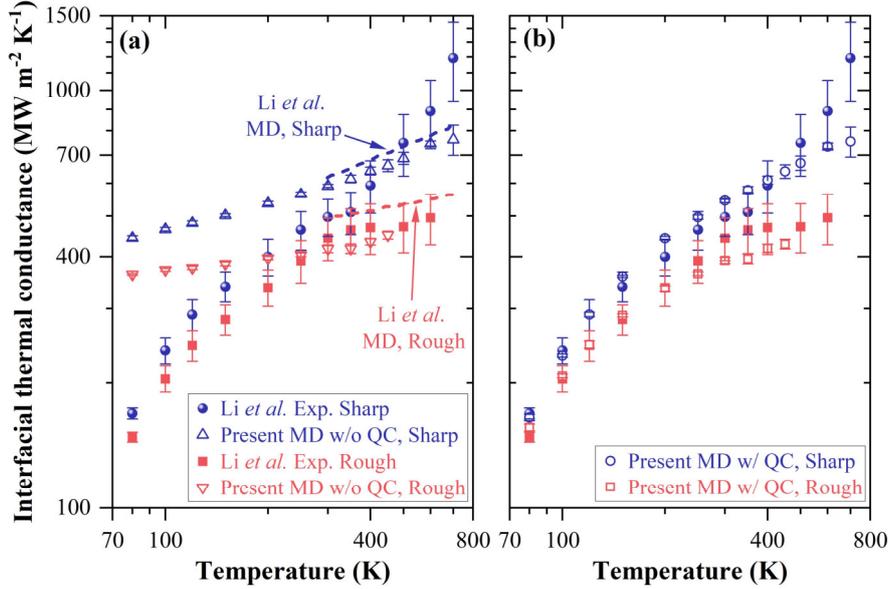

FIG. 2. The temperature-dependent interfacial thermal conductance of sharp and rough Si/Al interface configurations by NEMD simulations (a) without and (b) with quantum correction (QC) compared to the theoretical and experimental data by Li *et al.* [14]. Each ITC result by the present MD represents the average of six independent simulations.

The interfacial spectral thermal conductance ($G_{\omega 2}$) from $L_0$ to $R_0$ region, at two representative temperatures of 100 and 300 K, are shown in Fig. 3. The spectral thermal conductance at 100 K without quantum correction is significantly higher than that with quantum correction as shown in Fig. 3(a). This clearly demonstrates the inadequacy of the assumption in classical MD calculations that all phonon modes are equally excited at low temperatures [57]. Therefore, the quantum correction is included in the spectral thermal conductance analyses hereafter. Obviously, in the rough interface, the phonon spectral contribution above ~3.5 THz is considerably lower than that in the sharp configuration while is appreciably higher than the latter in the low-frequency range below ~3.5 THz. Note that a different trend has been obtained by harmonic NEGF method [14] where the transmissivity of phonons below 4 THz in rough Si/Al interface is obviously lower than that of sharp one. This indicates the importance of including lattice anharmonicity in studying the effect of interfacial roughness, which unexpectedly enhance the spectral contribution to ITC in the low-frequency range. At room temperature (300 K) as shown in Fig. 3(b), the quantum effect becomes weak suggesting that almost all phonon modes are excited. Moreover, the trend of spectral contribution at 300 K is similar to that of 100 K, except that the spectral contribution at the sharp interface is significantly higher than the rough one above ~3.5 THz. This suggests that phonon scattering induced by atomic disorder in interfacial region is much stronger than the ordered sharp configuration in moderate- and high-frequency range.

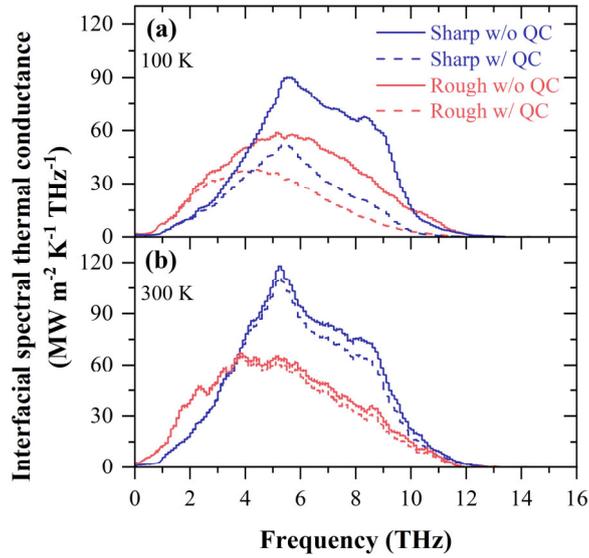

FIG. 3. Interfacial spectral thermal conductance via MD at (a) 100 K and (b) 300 K. Blue represents the sharp Si/Al interface configuration, while red represents the rough configuration. The solid lines denote the results without quantum corrections (QC), and the dashed lines indicate the results with quantum corrections.

Then, we focus on the contribution of phonon to thermal transport around the interface as shown in Fig. 4, which illustrates the spectral thermal conductance before ($G_{\omega 1}$), at ($G_{\omega 2}$), and after ($G_{\omega 3}$) the interface. At the sharp interface, phonons in the entire frequency range of crystal-Si and crystal-Al contribute to $G_{\omega 1}$ and $G_{\omega 3}$ before and after phonon crossing the interface, as shown in Fig. 4(a) and Fig. 4(c). In contrast, phonons with frequencies exceeding the cutoff frequency of crystal-Al (~10 THz) have contribution to the $G_{\omega 2}$ as shown in Fig. 4(b), which is relevant to the special interfacial modes as to be shown later by the local VDOS analysis. The strong variation of spectral thermal conductance suggests that phonon energy changes during the transport from the Si side to the Al side, as a clear signature of inelastic phonon scattering. From an overall perspective, the contribution of moderate-frequency phonons to thermal transport gradually increases with rising temperature whereas the high-frequency part does not vary too much, which explains the measured significant increase of ITC of the sharp interface above 300 K (shown in Fig. 2). At the rough interface, the frequency-dependent behaviors of $G_{\omega 1}$, $G_{\omega 2}$ and $G_{\omega 3}$ shown in Fig. 4(d)-(f) are generally similar to those in the sharp configuration, yet with appreciably smaller amplitudes in moderate- and high-frequency range due to the roughness scattering of phonons. However, the temperature dependence is quite different, *i.e.*, the spectral thermal conductance almost does not change when the temperature increases from 300 K to 450 K, which correlates well with the saturation of ITC above 300 K in rough interface (shown in Fig. 2).

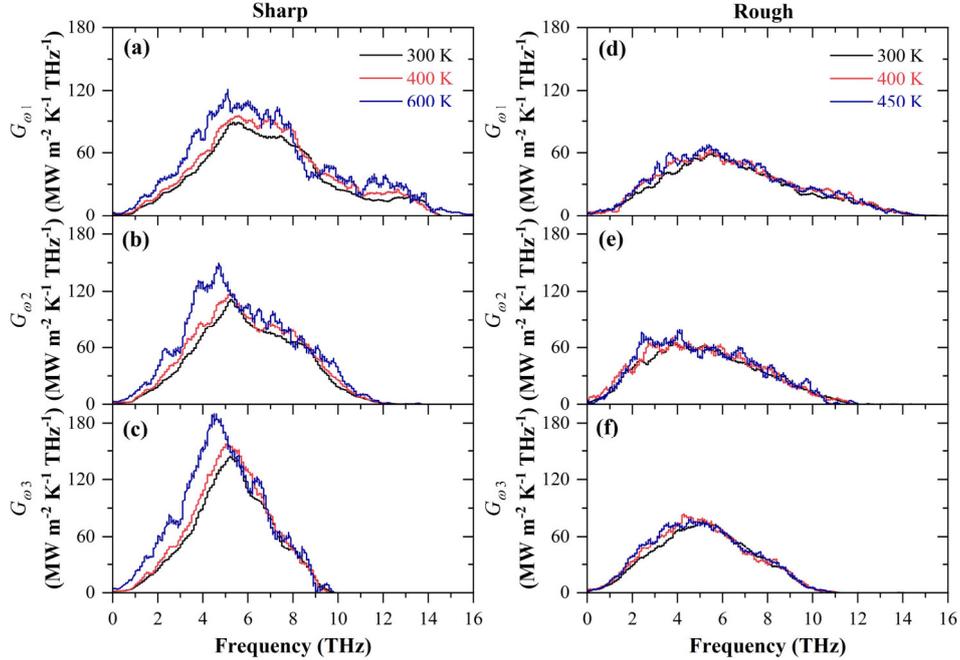

FIG. 4. Spectral thermal conductance with quantum correction at few temperatures around the sharp (a)-(c) and rough (d)-(f) Si/Al interfaces. $G_{\omega 1}$, $G_{\omega 2}$ and $G_{\omega 3}$ correspond separately to $\mathbf{J}_{\omega 1}$, $\mathbf{J}_{\omega 2}$ and $\mathbf{J}_{\omega 3}$ in Fig. 1(a).

*3.2 Spectral energy exchange around the interface*

The analysis in Sec. 3.1 only demonstrates that the interfacial roughness significantly influences the spectral thermal conductance around the interface, yet does not elucidate how it impacts inelastic phonon scattering. Here, we provide deeper analysis based on the spectral energy exchange $\delta G_{\omega 12}$ and $\delta G_{\omega 23}$, as described in Sec. 2, in $L_0$ and $R_0$ regions around the interface as shown in Fig. 5. At the sharp interface, Fig. 5(a) and Fig. 5(b) describe the net effect of inelastic scattering of phonons before and after crossing the interface, respectively. Specifically, the high-frequency phonons above ~9.5 THz are predominantly annihilated to generate phonons in the 2-6 THz range in $L_0$ region. Note that spectral energy exchange below ~2 THz is close to zero, suggesting that those phonons transport in an elastic way. In $R_0$ region, the phonons above ~7 THz are predominantly annihilated to generate phonons below 7 THz. Moreover, appreciable phonon generation can be observed below ~2 THz in $R_0$ region only at very high temperature (600 K) as seen Fig. 5(b). At the rough interface, the net effect of inelastic scattering in the $L_0$ region mainly involves phonon annihilation above ~9 THz to generate phonons below ~5 THz, as shown in Fig. 5(c) and (d). Interestingly, phonons below ~3 THz in the $L_0$ region also exhibit strong generation even at room temperature, which shall correspond to the larger spectral contribution below ~3.5 THz in rough configuration shown in Fig. 3. Compared to the $L_0$ region, phonons in the $R_0$ region experience inelastic scattering in a different way, where phonons both below ~4 THz and above ~8 THz are annihilated, generating phonons in the 4-8 THz range before entering the $R_1$ region (Al side), as seen in Fig. 5(d). With increasing temperature, the larger amplitude of spectral energy exchange at the sharp interface indicates stronger inelastic phonon scattering, while the rough interface shows a trend toward saturation. This contrast provides the explanation of the significant increase and saturation of ITC respectively for sharp and rough interfaces above 300 K seen in Fig. 2. In all, the interfacial roughness reduces the inelastic phonon scattering except a special increase in low-frequency range below ~3 THz, as to be further discussed later. Note that the spectral energy exchange can only reveal the overall net effect of inelastic phonon scattering, whereas the specific phonon scattering processes around the interface are still missing and pending to be dug out in the future. Nevertheless, it provides more information than that given by the classical SHC analysis [11,46,47], and thus will promote the analysis method in the framework of MD.

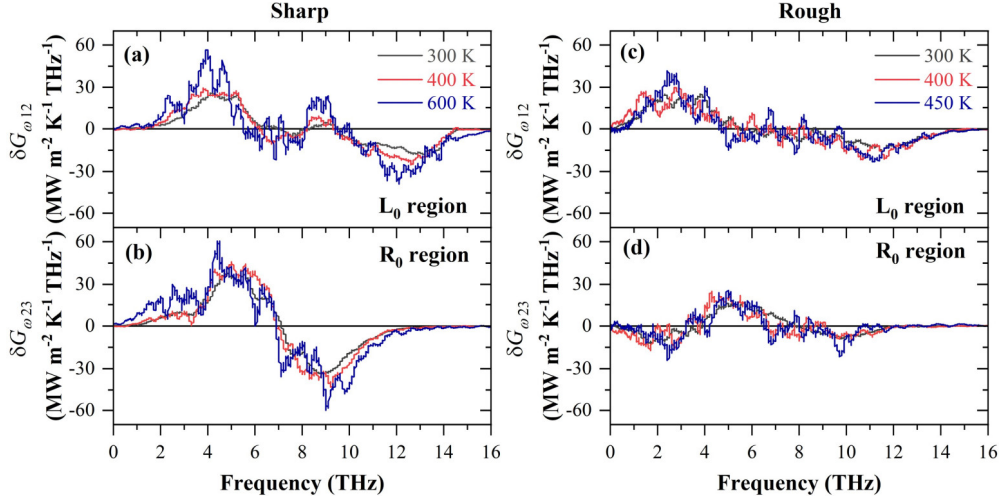

FIG. 5. Spectral energy exchange at few temperatures within the two nearby regions of (a)-(b) sharp and (c)-(d) rough Si/Al interfaces. The $\delta G_{\omega 12 (23)} > 0$ and $\delta G_{\omega 12 (23)} < 0$ indicate net phonon generation and annihilation, respectively, where $\delta G_{\omega 12} = \delta G_{\omega 2} - \delta G_{\omega 1}$ and $\delta G_{\omega 23} = \delta G_{\omega 3} - \delta G_{\omega 2}$.

Finally, we illustrate the variation of local VDOS around the Si/Al interface along the transport direction as presented in Fig. 6, as a support for the preceding results of spectral thermal conductance and spectral energy exchange. The local VDOS in the $L_3$ and $R_3$ region can be considered as the local VDOS of crystal-Si and crystal-Al, respectively. The existence of the interface alters the phonon states near the interface. In the sharp interface, although atoms in the $L_0$ region maintain the crystal-Si structure, the local VDOS in this region still differs somehow from that of crystal-Si, as shown in Fig. 6(a). Similarly, the local VDOS in the $R_0$ region deviates from that of crystal-Al. The small VDOS above 10 THz represents special interfacial modes, which contributes to the spectral contribution to ITC beyond the cut-off frequency of Al shown in Fig. 4(b). For the rough interface, the local VDOS in the disordered $L_0$ and $R_0$ regions near the interface differs much more significantly from that of the ordered crystal-Si and crystal-Al regions on both sides, as shown in Fig. 6(b). Obviously, the rough interface exhibits distinct interfacial phonon modes below ~3 THz, which provides the channels for appreciable phonon generation and annihilation below ~3 THz shown in Fig. 5(c) and (d). It has been also demonstrated in previous experimental studies that the local VDOS at interface is higher than that within the region of the parents' materials in the low frequency range [60,61]. In contrast, in regions further away from the interface, *i.e.*, $L_1$ and $R_1$ regions, the local VDOS closely matches that of crystal-Si and crystal-Al, respectively, indicating the short range of interfacial modes, as consistent with previous studies [6,7,10,44].

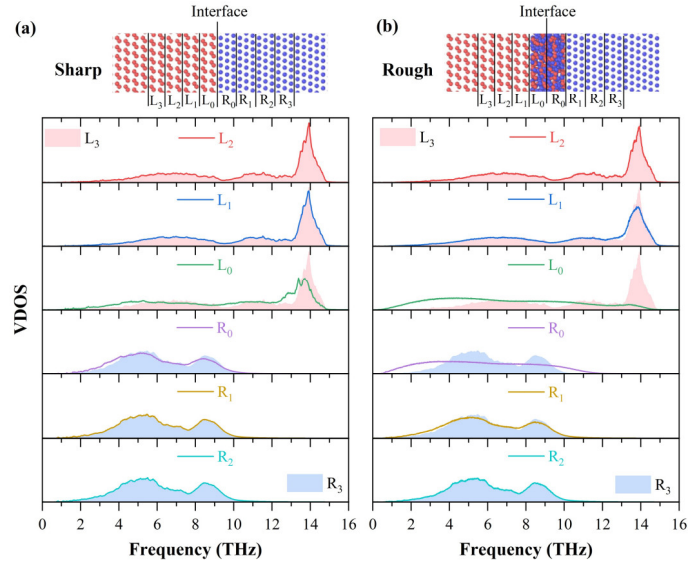

FIG. 6. The local vibrational density of states (VDOS) from MD around the (a) sharp and (b) rough interfaces. The length of each region from $L_3$ to $R_3$ is 0.7 nm.

## 4. Conclusions

In summary, we use non-equilibrium molecular dynamics simulation and introduce spectral energy exchange analysis to investigate interfacial thermal transport at both sharp and rough Si/Al interfaces. The predicted interfacial thermal conductance (ITC) results with quantum correction show a good agreement with experimental results in a very broad range of temperatures. The interfacial roughness generally reduces the spectral contribution of moderate- and high-frequency phonons to ITC, while enhances that of low-frequency phonons below ~3.5 THz. The spectral energy exchange analyses show more annihilation of high-frequency phonons and generation of moderate-frequency phonons around the sharp interface than that around the rough interface. However, we find unexpected stronger inelastic scattering of phonons below ~3 THz around the rough interface due to the unique interfacial phonon modes as evidenced by the local vibrational density of states. Our findings provide insights into the microscopic scattering mechanism of interfacial thermal transport. The present spectral energy exchange analysis will be a useful tool in the frame of MD to understand the inelastic phonon scattering at solid/solid interfaces.

**Acknowledgements**

J. Xu would like to thank Dr. Y. Xu (HKUST) for helpful discussions. Y. Guo would like to appreciate the financial support of the starting-up funding (AUGA2160500923) from Harbin Institute of Technology and the NSFC Fund for Excellent Young Scientists Fund Program (Overseas).

**Data availability**

The data that support the findings of this study are available from the corresponding author upon reasonable request.

**Appendix A. Phonon dispersion from empirical potentials**

As shown in Fig. 7, the ADP display in a good agreement with experimental data [62,63] for describing the phonon dispersion relationship of crystal-Si and crystal-Al compared to the MEAM potential. Hence, we chose the ADP to describe the atomic interactions between Al-Al, Si-Si and Al-Si.

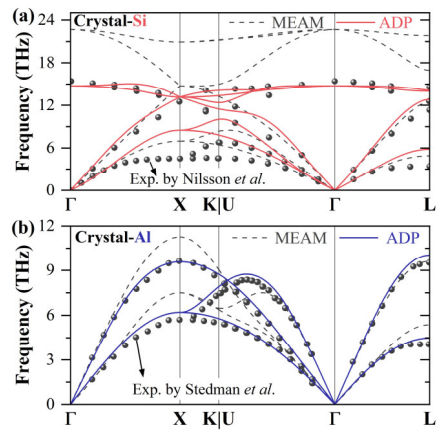

FIG. 7. Calculated phonon dispersion curves of (a) crystal-Si and (b) crystal-Al using MEAM (dash line) and ADP (solid line) compared to experimental measurements (solid dot) [62,63].

**Appendix B. Size-dependence test in MD for ITC calculation**

After the structure relaxation using ADP, the lattice constants of Si and Al unit-cell along [111] direction are 3.867 and 2.850 Å at 300 K, respectively, which leads to a lattice mismatch of the cross section. Since the model slice along the [111] plane is rhombic, we transform it into a rectangular cell for the convenience of MD simulations, and then expand it into a square cell. To test the effect of interfacial strain, ~3.9 nm$^2$ (5 uc × 5 uc for Si [111] and 7 uc × 7 uc for Al [111]), ~15.6 nm$^2$ (10 uc × 10 uc for Si [111] and 14 uc × 14 uc for Al [111]) and ~64.6 nm$^2$ (21 uc × 21 uc for Si [111] and 28 uc × 28 uc for Al [111]) are used as cross-sectional areas of the simulation models. Results show that the difference of ITC using these three simulation sizes is small as shown in Fig. 8(a), which indicates that the stress and strain effect on our simulation results can be ignored in our NEMD simulations. Due to the massive computational cost using ADP, we consider a thermal transport length of 20 nm, which is sufficient as shown in Fig. 8(b).

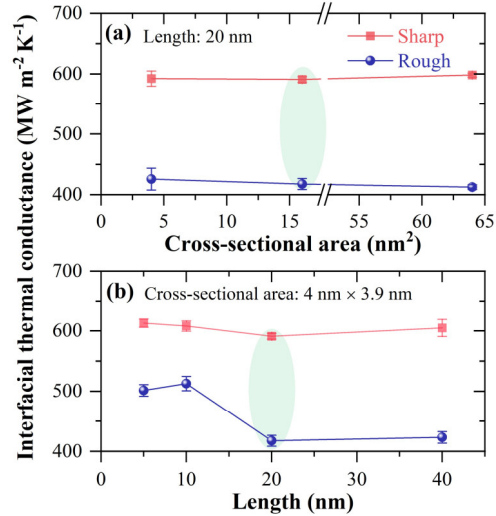

FIG. 8. Calculated ITC values as a function of (a) cross-sectional area and (b) transport length at 300 K. Each ITC result represents the average of six independent simulations and the error-bar is small.